\documentclass{article}
\usepackage{graphicx,subfigure,epstopdf}
\title{\bf Advantage of a lead swimmer in drafting}
\author{J. Westerweel, K. Aslan, P. Pennings \& B. Yilmaz \\[1ex]
{\small Laboratory for Aero- \& Hydrodynamics, Delft University of Technology} \\ 
{\small Mekelweg 2, 2526 CD Delft, The Netherlands}}

\begin{document}
\maketitle

\begin{abstract}
We present results from model tests to investigate the effect of drafting in swimming, in particular for the lead swimmer. The drag for  scaled-model passive swimmers was determined accurately at Froude numbers comparable to conditions for actual human swimmers. Several positions of a draft swimmer at different separations behind and alongside the lead swimmer were investigated. It was found that a lead swimmer can experience an advantage from a draft swimmer. Several other positions of the draft swimmer relative to the frontal wave generated by the lead swimmer were also considered. These results indicate favourable and undesirable positions during passing. 
\end{abstract}

\section{Introduction}
\label{sec:introduction}
It is well known that drafting has significant advantages in reducing the drag of multiple moving objects in air and water. Drafting is observed in nature in fish and birds \cite{Fish1995,McNeill2004}, and in sports drafting is common in ice skating, swimming and cycling \cite{Rundell1996, Chollet2000, Hausswirth2001, Chatard2003, Silva2008, Janssen2009, Blocken2013}. When drafting is done in pairs, it is evident that the draft position is advantageous, and that less flow drag is experienced. 
%%% It is not well known that 
Drafting also implies an advantage to the leading position, yet at a smaller degree in comparison to the draft position. The lead position benefits from an increase of the pressure in its wake due to the presence of the draft object, which results in a reduction of the total drag experienced in the lead position \cite{Blocken2013,Blocken2015}. Hence, the fact that in drafting a reduction of flow drag is achieved for \emph{both} the leading position and the draft position forms the basis for cooperation, both in nature and in sports.

The mutual benefits in drafting have been studied for cycling and skating. One can expect similar benefits for drafting in swimming, such as open water swimming and triathlon swimming, but the conditions for swimming are far more complicated than for cycling and skating. The largest complication in swimming is that both fluid motion and surface waves are generated \cite{Vennell2006}.

In previous studies \cite{Chatard2003,Silva2008,Janssen2009} the benefits of drafting were investigated for a draft swimmer, both directly behind the front swimmer as well as swimming to the side of the front swimmer. From these studies it was concluded that the optimal position for a draft swimmer is between 0.0 and 0.5 m back from the toes of the lead swimmer, resulting in a reduction of oxygen uptake of 11\% for the draft swimmer, in the case of an active lead swimmer. It was found in the same study that swimming alongside a lead swimmer did not provide any drafting advantages.

In this paper we present new results, based on laboratory experiments using scale models that drafting for swimming also has an advantage to the lead swimmer.
The advantage of using models is that very accurate and repeatable measurements can be made, while field tests are often compromised by fatigue of the swimmers during repeated trials and differences in fitness of the particular swimmers that participate in such tests.
\begin{figure}
	\begin{center}
		\includegraphics[width=0.8\columnwidth]{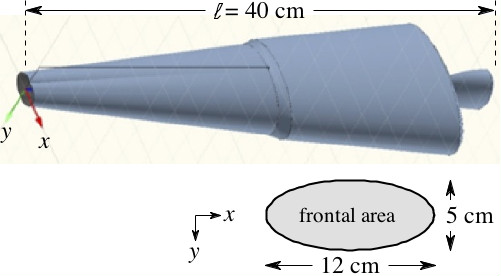}
	\end{center}
	\caption{Computer-generated isometric view of the model of a (passive) swimmer with the arms alongside its body. The model is based on an anthropomorphic data base ({\tt www.dined.nl}). Red and green arrows indicate lateral and vertical coordinate axes, respectively.}
	\label{fig:model}
\end{figure}

\section{Materials and methods}
\subsection{\em scaled swimmer models}
We performed measurements on scale models that represent (passive) swimmers.
The models were used to perform measurements in various positions.
The models resemble a human with the arms alongside its body, as shown in Figure~\ref{fig:model}. The shape of the model is based on an anthropomorphic data base (see: {\tt www.dined.nl}). 
%%%
We preferred to use a standardized model, rather than more realistically looking mannequin models that may not be standardized. Like the more detailed model used by Marinho {\em et al.} \cite{Marinho2009d}, the dined model is considered to have its arms alongside the body. Although more sophisticated models exists, e.g. \cite{Nakashima2012}, the proposed dined model would suffice for the objectives of this study.
%%%

The length of each model is 0.40 m. The frontal area $A$ has the shape of an ellipse with a long axis of 0.12 m and a short axis of 0.05 m, which gives $A$ = 47 cm$^2$.
The models are made out of 
%%%
polyurethane foam (which has a density of 180 kg/m$^3$)
%%%
on a \emph{computer numerical control} (CNC) machine, and later a thin lead cover was added to give the model the proper buoyancy and position in the water 
%%%
that mimics the buoyancy and position of actual swimmers.

\begin{figure}
	\begin{center}
		\includegraphics[width=0.7\columnwidth]{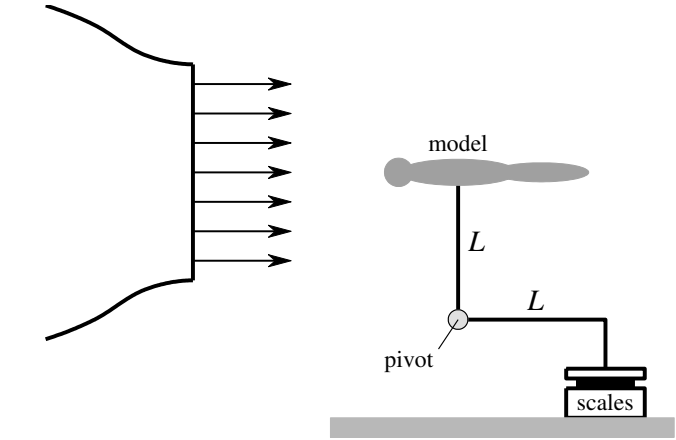}
	\end{center}
	\caption{Schematic of the wind tunnel measurements of the fully-submersed drag force of the model swimmer, using a scale to accurately measure the drag force. The cross section of the open wind tunnel at the contraction exit is 0.4$\times$0.4~m$^2$ and the length of the model is 0.4 m.}
	\label{fig:windtunnel}
\end{figure}

In order to represent a flow condition that can be considered to be representative for an actual swimmer, including the effect of the waves, it is important that the model situation matches the Froude number of an actual swimmer \cite{Naemi2010}. The Froude number $Fr$ is defined as
\begin{equation}
	Fr = \frac{U}{\sqrt{g \ell}},
\label{eq:def_froude_nr}
\end{equation}
where $U$ is the velocity, $g$ the gravitational acceleration, and $\ell$ a characteristic length scale. In this study we take for $\ell$ the length of the model. The Froude number for a trained triathlete is between 0.23 and 0.35 \cite{Vennell2006}, given that $U$ =  1.0--1.6 m/s, $\ell$ = 2 m, and $g$ = 9.81 m/s$^2$.
The drag coefficient $C_D$ is defined as
\begin{equation}
	C_D = \frac{D}{\frac12\rho U^2 A},
\label{eq:cd}
\end{equation}
where $D$ is the drag force, $\rho$ the fluid density, and $A$ the frontal area. 
% Note that we assume here that the model would be fully submerged.
%
For a proper dynamic scaling of the measured drag coefficient it is important that the model flow condition also matches the same Reynolds number $Re = U\ell/\nu$, where $\nu$ is the kinematic viscosity of the fluid. Given that $\ell$ appears as a square root in the denominator of $Fr$ and in the numerator of $Re$, it is evident that the dynamic similarity conditions for both $Fr$ and $Re$ cannot be met (for the same fluid and gravitational acceleration). However, for high Reynolds number (larger than 10$^5$) the drag coefficient only weakly depends on $Re$, which relaxes the requirement for matching the exact Reynolds number, while at Froude numbers between 0.2 and 0.3 the wave drag dominates the total drag and strongly depends on the value of $Fr$ \cite{Vennell2006}. This is why scaling based on the Froude number was chosen to be prevalent in this study.

\begin{figure}[h]
	\begin{center}
		\includegraphics[width=0.7\columnwidth]{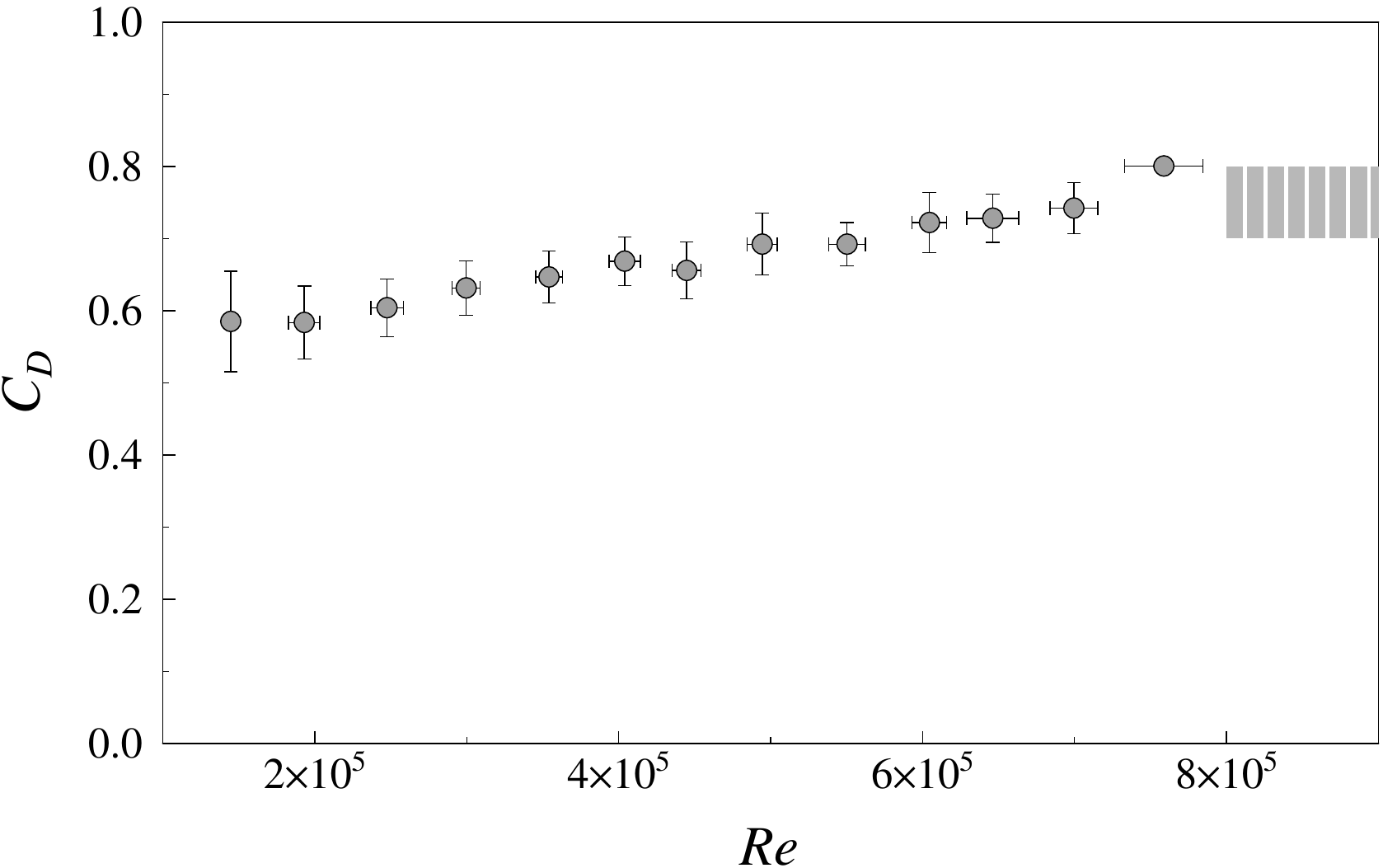}
	\end{center}
	\caption{The measured drag coefficient $C_D$ of the model as a function of the Reynolds number $Re$. The shaded region represents the drag coefficient estimated from a numerical simulation for $Re$ = 3--5$\times$10$^6$ \cite{Marinho2009b}, which is outside the range of the graph. Vertical error bars represent the 95\% reliability interval for the drag coefficient; horizontal error bars the uncertainty in wind tunnel speed. Error bars smaller than the size of the symbols have been omitted.}
	\label{fig:cd}
\end{figure}

In order to evaluate that our model would be representative of an actual passive swimmer, we took measurements of the drag force in the case of a fully submersed model. These results could then be compared against those of a numerical study by \cite{Marinho2009b}. 
Hence, the drag coefficient for a fully submerged model as a function of the Reynolds number was determined by placing the model in a small open wind tunnel. The wind tunnel has a cross section of 0.4$\times$0.4~m$^2$. The total drag force was measured using a balance, which is shown schematically in Figure~\ref{fig:windtunnel}. The accuracy of the balance was verified by measurements on the drag force on a solid sphere.
Figure~\ref{fig:cd} shows the measured drag coefficients for the model in a wind tunnel. The largest Reynolds number that could be realized in the measurement was around 8$\times$10$^5$ (which would correspond to an actual swim speed of around 0.4 m/s).
The drag coefficient is not exactly constant, but shows a mild increase as a function of $Re$. For large $Re$ the drag coefficient approaches a value of 0.7--0.8, which was found from a numerical simulation for a submerged passive swimmer with its arms alongside its body \cite{Marinho2009b}. A lower drag coefficient of 0.4--0.5 would be found for a submerged passive swimmer with its arms extended forward. The slight increase of the drag coefficient with speed (viz., Reynolds number) was also observed by Vennell \emph{et al.} \cite{Vennell2006}, yet with lower absolute values since a passive submerged swimmer with arms extended forward was considered. 
\\

It should be noted that the wind tunnel measurements are not representative for the drag of a partially submerged model swimmer. When the models are partially submerged, the total drag is determined by shape drag, friction drag, and wave drag, of which the last one will be dominant while the shape drag and friction drag will be significantly reduced for a partially submerged model. However, to make the validation independent of a possibly ill-defined degree of submersion, we only considered here a fully submerged model to allow for a comparison with the results of Marinho \emph{et al.} \cite{Marinho2009b}. Hence, the value reported above only serves the purpose of validation. 

\begin{figure}
	\begin{center}
		\includegraphics[width=\columnwidth]{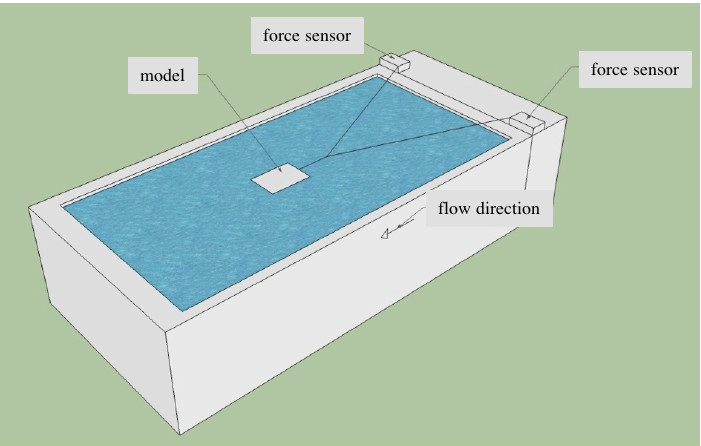}
	\end{center}
	\caption{Schematic of the flume for the model tests. The length of the flume is 3~m, and its width is 1.5 m. The water depth is 0.3 m.}
	\label{fig:waterchannel}
\end{figure}

\subsection{\em laboratory measurements}
\begin{figure}
	\begin{center}
		\subfigure[~]{\includegraphics[width=0.47\columnwidth]{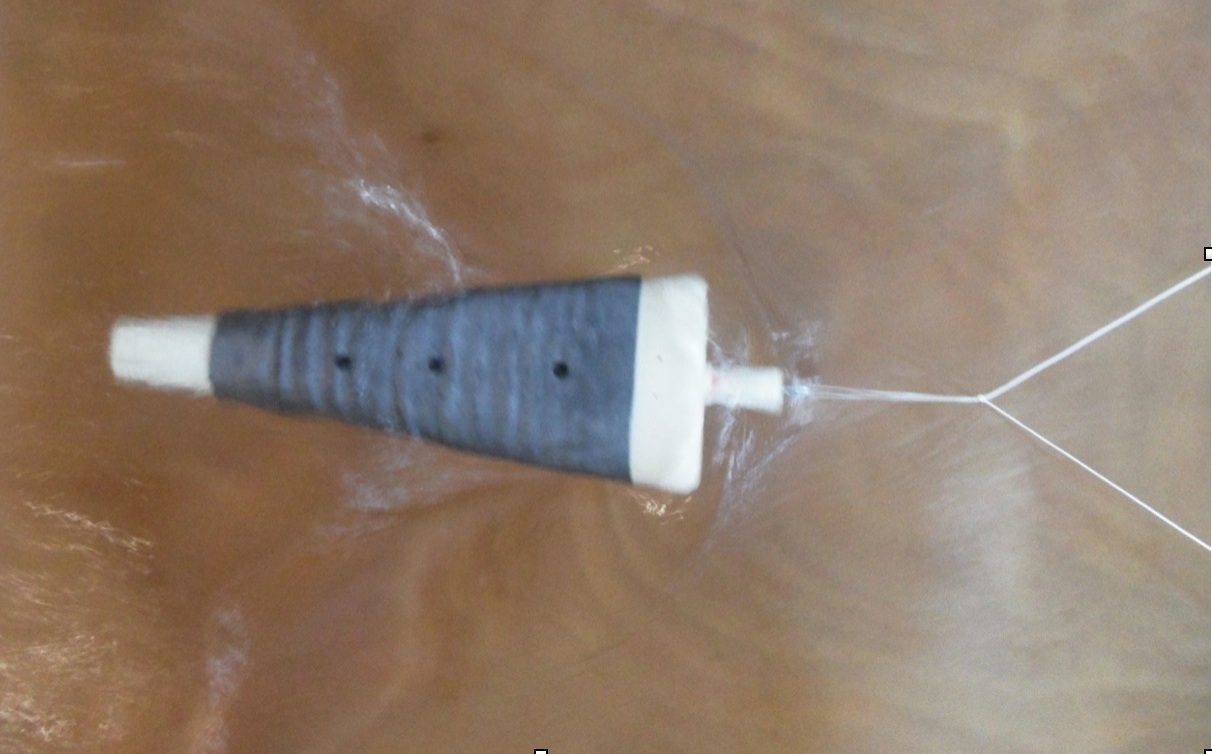}}
		\hfill
		\subfigure[~]{\includegraphics[width=0.47\columnwidth]{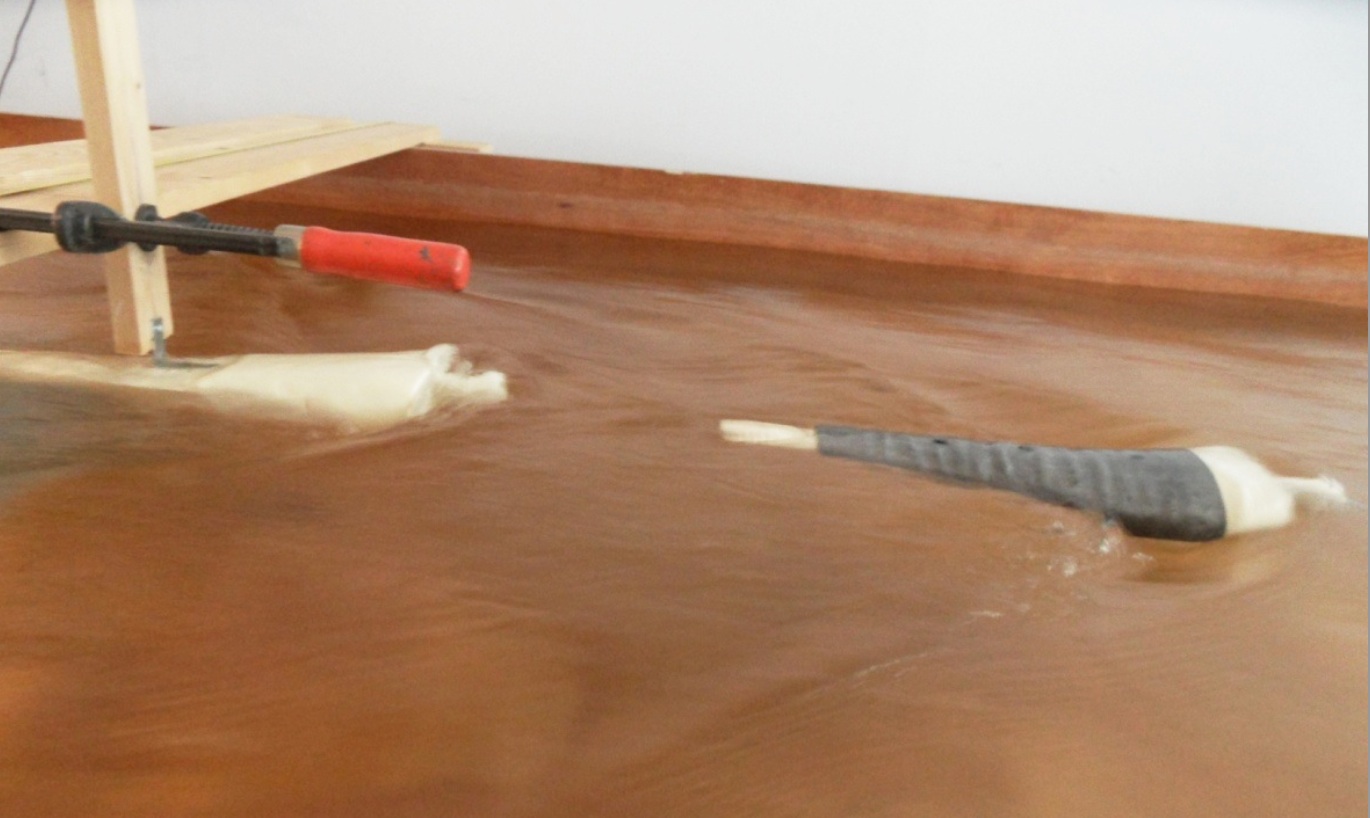}}
	\end{center}
	\caption{(a) Top view of the model in the flume at a flow velocity of $U$ = 0.55 m/s. The length of the model is 0.40 m (see Fig.~\ref{fig:model}).
	(b) Photograph of two models in a drafting configuration. Here the drag force is measured on the model lead swimmer, while the position of the model draft swimmer is fixed. The bow wave generated by the model lead swimmer is clearly visible.
	}
	\label{fig:photo_single_model}
\end{figure}

The drag of the model swimmer was measured in a small flume, shown schematically in Figure~\ref{fig:waterchannel}. 
%%% hier beschrijving van het waterkanaal
The flume has a width of 1.5 m with a water depth of 0.3 m, and a length of 3 m.
%The flow speed was adjusted to match the Froude number of an actual swimmer; 
For the measurements we used a flow velocity of 0.55 m/s, which gives a Froude number of $Fr$ = 0.28. This is at the center of the Froude-number range mentioned above, and corresponds to a human swim speed of 1.23 m/s (for $\ell$ = 2 m).
Figure~\ref{fig:photo_single_model}a shows a top view of the model in the flume. 
The drag force, which is now a combination of friction drag, form drag and wave drag, is measured by means of two force sensors (Scaime EP2). The model is connected with two strings to each of the force sensors, as shown in Figure~\ref{fig:waterchannel}. This stabilizes the position of the model and avoids an upstream disturbance in the water flow.
An equal reading on both sensors indicates that there is no lateral force on the model, i.e. the measured force is the drag force in the direction of the fluid motion. All measurement were performed in the center of the flume in order to minimize any possible effect from the side walls and any dependence on the downstream position.

%\begin{figure}
%	\begin{center}
%		\includegraphics[width=\columnwidth]{photo_two_models.jpg}
%	\end{center}
%	\caption{}
%	\label{fig:photo_two_models}
%\end{figure}

\begin{figure}[t]
	\begin{center}
		\includegraphics[width=0.6\columnwidth]{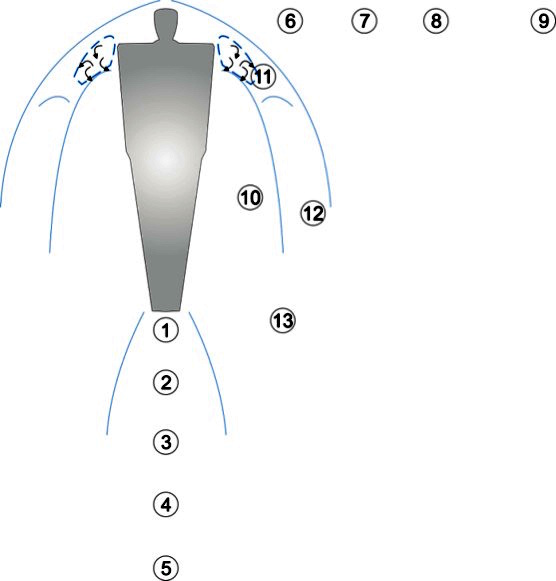}
	\end{center}
	\caption{The `head' positions of the draft model swimmer relative to the lead model swimmer. The curved lines indicate the approximate positions of the `bow' waves (see also Figure \ref{fig:photo_single_model}b).}
	\label{fig:positions}
\end{figure}

First, as a reference, the drag force on a single model swimmer was measured. We then placed a second identical model at various positions behind and alongside of the lead swimmer model, and  measured also the force again on the lead swimmer model. An overview of the different positions is given in Figure~\ref{fig:positions}.
%
%%% hoe zijn de metingen uitgevoerd, herhalingen, nauwkeurigheid
The drag force was measured over a duration of 10 seconds at a sampling rate of 100~Hz using a signal conditioner (Scaime CPJ) and a data acquisition board (LabView). Each measurement was repeated 10 times. This provided adequate accuracy to determine the drag force with an accuracy of about 1\%.

The two model swimmers were arranged in the positions as indicated in Figure~\ref{fig:positions}: positions 1--5 are with the draft swimmer behind the lead swimmer, at various separations. 
We then considered the drag for the two model swimmers side-by-side (viz., lateral drafting), also at increasing separations (positions 6--9). 
To further explore the effect of drafting for different relative positions in relation to the `bow' wave, we considered the positions 10--13, as shown in Figure~\ref{fig:positions}. These would correspond to: a position where the head of the draft swimmer is near the hip of the lead swimmer (position 10); a position where the head of the draft swimmer is near the shoulder of the lead swimmer (position 11); a position where the head of the draft swimmer is near the hip of the lead swimmer, but in the bow wave of the lead swimmer (position 12); and where the head of the draft swimmer is at the feet of the lead swimmer, but laterally offset so that its shoulders are next to the feet of the lead swimmer (position 13). These positions are relevant when a draft swimmer tries to pass a lead swimmer. It is noted that position 10 is close to the optimal position recommended by Chatard \& Wilson \cite{Chatard2003}.

\begin{table}[t]
	\caption{The drag forces for the lead ($D_1$) and draft ($D_2$) model swimmers relative to the drag force $D_0$ of a single model swimmer. For positions 6--9, $D_1=D_2$.}
	\label{tab:results}
	\begin{center}
		\begin{tabular}{cccc}
		\hline\hline
		position &relative  &lead          &draft \\
	         		      &separation &swimmer &swimmer \\
	 		      &$s / \ell$ &$D_1/D_0$ &$D_2/D_0$ \\
		\hline
	 	1 &0.00 &0.79 &0.36 \\
	 	2 &0.13 &0.88 &0.47 \\
	 	3 &0.25 &0.94 &0.43 \\
	 	4 &0.50 &0.94 &0.43 \\
	 	5 &0.75 &0.99 &0.48 \\
		\hline
		6 &0.00 &\multicolumn{2}{c}{1.55} \\
		7 &0.13 &\multicolumn{2}{c}{1.24} \\
		8 &0.25 &\multicolumn{2}{c}{1.19} \\
		9 &0.50 &\multicolumn{2}{c}{1.11} \\
		\hline
		10 & &0.95 &0.48 \\
		11 & &0.88 &1.02 \\
		12 & &1.01 &0.88 \\
		13 & &0.92 &0.75 \\
		\hline\hline
		\end{tabular}
	\end{center}	
\end{table}

\section{Results}
The mean drag force $D_0$ for the single model swimmer was 0.439$\pm$0.003~N (P=0.95). This gives a drag coefficient according to (\ref{eq:cd}) of $C_D$ = 0.62. This is in good agreement with the result of Vennell \emph{et al.} \cite[Fig. 4]{Vennell2006} for the same Froude number of a passive swimmer at the water surface. 
The measured relative variations of the drag forces in different drafting configurations, as shown in Figure~\ref{fig:positions}, are given in Table~\ref{tab:results}. We first consider the positions 1--5, where the draft swimmer is behind the lead swimmer. The results show that the draft swimmer experiences a significant reduction in drag. The largest reduction is found directly behind the lead swimmer, but the advantage remains significant over larger separations, exceeding that of one full body length. These results confirm earlier measurements on the advantage of drafting for a draft swimmer \cite{Janssen2009}. 
%%%
In this position, the drag for the draft swimmer is between 45\% and 53\% of the drag for the lead swimmer, which is close to the value of 56\% found by Silva \emph{et al.} \cite{Silva2008} using a computational approach. 
%%%
The present measurements also show that the lead swimmer also experiences a drag reduction from the presence of a draft swimmer. The effect is more modest than for the draft swimmer, and the advantage has disappeared when the separation of the draft swimmer is more than 50--75\% of the body length.

In a second series of measurements we considered side-by-side swimming at different separations; see positions 6--9 in Figure~\ref{fig:positions}. In this case, both swimmers experience the same drag. Evidently, swimming close together significantly increases the drag for both swimmers, as a result of the `bow' wave interaction. At a separation of more than 50\% of the body length (position 9), the increase in drag has become small, and swimmers can be considered as individual swimmers.
It should be mentioned here that in these positions we noted that the two force sensors gave unequal  readings, which is indicative of a lateral force towards the opposite swimmer. 
This lateral force may lead to interference and eventual collision of two actual swimmers during passing when they are swimming too close together.
\\

\section{Discussion}
In the case where the swimmers are directly behind each other, the draft swimmer experiences the largest advantage. This is in partial agreement with previous measurements: with passive swimmers we found that the reduction in drag is between 50 and 60\% for the draft swimmer, while Janssen \emph{et al.} \cite{Janssen2009} measured a reduction of 21\% for the draft swimmer in the case of active swimming. 

The present model measurements also indicate that the lead swimmer in drafting also experiences an advantage, albeit at a smaller degree than for the draft swimmer. Given that the advantage for a draft swimmer with active swimming is not as large as for passive swimmers, one may expect that the advantage for the lead swimmer in active swimming may not be as large as found in these measurements. 
\\

The results show that a draft swimmer should definitely avoid position 11, which is the least favorable position when passing a lead swimmer. Interestingly enough, this position is most favorable to the lead swimmer. Hence, this can be used by a lead swimmer to impede the passing of a faster swimmer. It should be noted here that in actual swimming this position may not occur very frequently, as moving arms may interfere or even collide in this position. The present measurements confirm that position 10 is the optimal lateral drafting position, as recommended by Chatard \& Wilson \cite{Chatard2003}. However, the present measurements indicate that besides the advantage to the draft swimmer, there also appears to be an advantage to the lead swimmer for this position.

\section{Conclusion}
The objective of this study was to determine whether the lead swimmer in drafting experiences an advantage from the presence of a draft swimmer. In order to achieve accurate measurement results, scale-model tests were conducted in a water flume. The models were designed to have the same Froude number as actual (passive) swimmers. For the dynamic scaling we matched the Froude number,  since the wave drag is dominant. The measurements on an individual model swimmer indicated that the total drag force was comparable to results from computational fluid dynamics results and full-scale towing models. Hence, the scaled model study is considered to provide reliable results for drafting studies with passive swimmers.
\\

The laboratory measurements with scaled passive model swimmers were used to determine favorable and unfavorable positions for drafting in swimming through the direct measurement of the mean drag for various drafting configurations. These measurements confirm the advantages for the draft swimmer; they also indicate that drafting in swimming would imply an advantage to the lead swimmer as well.
The model tests also suggest how to pass and how to impede being passed during swimming. 
\\

This study demonstrates how model studies can be performed to obtain accurate results that may form the basis for swimming strategies in open water, including triathlon swimming. The model tests can be accurately performed and provide repeatable results, as opposed to field test, where one has to deal with individual athletes and with indirect indicators, such as heart rate, to assess possible drafting advantages.
Evidently, the present laboratory results are only indicative and need to be validated in actual field trials. It is therefore recommended that extensive field trials are carried out to validate the model results. 
Another aspect that needs further investigation is the observation of a (small) lateral force in side-by-side swimming, which could lead to interference and collision of actual swimmers while swimming side-by-side or during passing.

\section*{Acknowledgement}
%The authors would like to thank prof. Peter Beek of the VU University, Amsterdam for his contributions and recommendation regarding this work.
Tarik Sezer and Ishak Guclu contributed to the experiments.
This work was in part supported by Grant 12186 `Optimization of propulsion in swimming and rowing' of the Technology Foundation STW.

\end{document}